\documentclass[12pt]{article}

\usepackage{epsfig}
\usepackage{amsmath}
\usepackage{amssymb}
\usepackage[round]{natbib}

\def\beq{\begin{equation}}
\def\eeq#1{\label{#1}\end{equation}}
\def\eeqn{\end{equation}}

\def\beqa{\begin{eqnarray}}
\def\eeqa#1{\label{#1}\end{eqnarray}}
\def\eeqan{\end{eqnarray}}

\let\bar=\overbar

\def\Dslash{\not{\hbox{\kern-4pt $D$}}}
\def\dslash{\not{\hbox{\kern-2pt $\del$}}}

\def\msb{{\bar{\ssstyle M \kern -1pt S}}}

\usepackage{fancyhdr,graphicx}
\def\Title#1{\begin{center} {\Large {\bf #1} } \end{center}}

\begin{document}
\bibliographystyle{plainnat}

\Title{Exploring the QCD phase transition in core collapse supernova
         simulations in spherical symmetry}

\bigskip\bigskip


\centerline{
T. Fischer\(^{\dagger}\),
I. Sagert\(^{\ddagger}\),
M. Hempel\(^{\ddagger}\),
G. Pagliara\(^{\ddagger}\),
J. Schaffner-Bielich\(^{\ddagger}\),}
\centerline{
A. Mezzacappa\(^*\),
F.-K. Thielemann\(^{\dagger}\)
and
M. Liebend\"orfer\(^{\dagger}\)}

\bigskip\bigskip

\begin{raggedright}
\(^{\dagger}\)Department of Physics, University of Basel, Klingelbergstrasse 82,
4056 Basel, Switzerland\\
\(^{\ddagger}\) Institut f\"ur Theoretische Physik,
Ruprecht-Karls-Universit\"at,
Philosophenweg 16,  69120 Heidelberg, Germany\\
\(^*\) Physics Division, Oak Ridge National Laboratory,
Oak Ridge, Tennessee 37831-1200, US
\bigskip\bigskip
\end{raggedright}

\begin{abstract}
For finite chemical potential effective models of QCD predict
a first order phase transition.
In favour for the search of such a phase transition in nature, we construct
an equation of state for strange quark matter based on the MIT bag model.
We apply this equation of state to highly asymmetric core collapse
supernova matter with finite temperatures and large baryon densities.
The phase transition is constructed using the general Gibbs conditions,
which results in an extended coexistence region between the pure hadronic
and pure quark phases in the phase diagram, i.e. the mixed phase.
The supernovae are simulated via general relativistic radiation
hydrodynamics based on three flavor Boltzmann neutrino transport
in spherical symmetry.
During the dynamical evolution temperatures above \(10\) MeV,
baryon densities above nuclear saturation density and
a proton-to-baryon ratio below \(0.2\) are obtained.
At these conditions the phase transition is triggered which leads
to a significant softening of the EoS for matter in the mixed phase.
As a direct consequence of the stiffening of the EoS again for matter
in the pure quark phase, a shock wave forms at the boundary between
the mixed and the pure hadronic phases.
This shock is accelerated and propagates outward which releases a burst
of neutrinos dominated by electron anti-neutrinos due to the lifted
degeneracy of the shock-heated hadronic material.
We discuss the radiation-hydrodynamic evolution of the phase transition
at the example of several low and intermediate mass Fe-core progenitor
stars and illustrate the expected neutrino signal from the phase transition.
\end{abstract}

\section{Introduction}

The investigation of the QCD phasediagram via heavy-ion experiments at BNL's
RHIC, CERN's LHC and the FAIR facility at GSI poised to explore the properties
of QCD matter under extreme conditions.
Three of the most important aspects are the behaviour and the position
of the critical points in the phase diagram, the phase transition from
hadronic matter to quark matter at finite chemical potentials
and the properties of the predicted quark phases.
In search for these aspects, observations of astronomical objects and
astrophysical processes that are assumed to contain quark matter could
be helpful.
In such favour, Cold and isolated or binary neutrons stars (NS) have long
been served as powerful objects to probe the equation of state (EoS)
for hadronic as well as for quark matter.
In the latter case the NS is called a hybrid star
if in addition to the quark core a hadronic envelope is present.
The astrophysical processes that leave an isolated NS
as the final remnant are core collapse supernovae
of low and intermediate mass Fe-core progenitor stars.
Naturally the question rises at which stage
during the dynamical evolution from a collapsing Fe-core
to an isolated NS the thermodynamic conditions are such
that the appearance of quark matter is favoured?
Even further in the case of a phase transition from hadronic matter
to quark matter, which hydrodynamic evolution can be expected
and is there a relation to observations?
The two intrinsically different scenarios are either
during the early post bounce phase when the temperatures
are moderately high
or during the cooling of the remnant, where deleptonisation
causes a temperature decrease and a density increase.
The present article discusses the first case due to the
critical conditions given by the quark matter EoS,
where a relation to the explosion mechanism is explored.
Therefore, a detailed study of core collapse supernovae including
radiation-hydrodynamics with spectral neutrino transport and
a sophisticated EoS for hot and dense asymmetric matter
is required to simulate the matter conditions accurately.

The first study of the QCD phase transition in
core collapse supernovae was published by \citet{TakaharaSato:1988},
suggesting a relation of the multi-peak neutrino signal
from supernova 1987A (see \citet{Hirata:etal:1988})
to the appearance of quark matter.
Using general relativistic hydrodynamics, they modelled the phase
transition via a polytropic EoS.
Due to the absence of neutrino transport
they could neither confirm nor exclude the suggested prediction
of a neutrino signal from the phase transition.
Additional microphysics was included in the study by
\citet{Gentile:etal:1993}, where general relativistic hydrodynamics
is coupled to a description of deleptonisation
during the collapse phase of a progenitor star.
Applying the MIT-bag model for the description of the quark phase,
they find the formation of a (second) shock wave
as a direct consequence of an extended
co-existence region between the hadronic phase and
the quark phase with a significantly smaller adiabatic index.
The second shock wave follows and merges with the first shock
from the Fe-core bounce after a few milliseconds.
However, due to the lack of neutrino transport
in the post bounce phase,
they were also not able to confirm the predictions made
for a possible neutrino signal from the phase transition.
The recent investigation by
\citet{Nakazato:etal:2008}
is based on general relativistic radiation hydrodynamics
with spectral three flavour Boltzmann neutrino transport.
They investigate very massive progenitors
of \(\sim 100\) M\(_\odot\) which collapse to a black hole.
Applying the MIT-bag model for quark matter,
the time after bounce for black hole formation is shortened
and corresponds to the appearance of quark matter,
where the central mass exceeds the maximum stable mass
given by the quark EoS.

We follow a similar approach and apply the MIT-bag model
for the description of quark matter in
general relativistic radiation hydrodynamics simulations,
based on spectral three flavor Boltzmann neutrino transport.
Our simulations are launched from low and intermediate mass progenitors,
where no explosions could be obtained in spherical symmetry.
We investigate the dynamical effects and discuss the possibility
of an observable in the neutrino signal related to the
QCD phase transition.

The manuscript is organised as follows.
In \S 2 we will present the standard core collapse supernova
scenario and in \S 3 we will lay down our
neutrino radiation hydrodynamics model
including both the hadron and quark EoSs.
In \S4 we will discuss the appearance of quark matter during the
early post bounce phase of core collapse supernovae of
intermediate mass Fe-core progenitors
and summarise the results in \S 5.

\section{Core collapse supernova phenomenology}

The Fe-core of massive progenitor stars in the mass range of 
\(9-75\) M\(_\odot\) collapse at the final stage of nuclear burning
due to photodisintegration and deleptonisation,
which reduces the proton-to-baryon ratio given by
the electron fraction \(Y_e\).
During the collapse, the density and temperature increase.
The collapse continues and at about \(\rho\simeq10^{13}\) g/cm\(^3\)
neutrinos are trapped.
At nuclear densities of \(\rho\simeq2-4\times10^{14}\) g/cm\(^3\),
the collapse halts and the central core is highly deleptonised
where \(Y_e\lesssim 0.3\).
The core bounces back and a shock wave forms which travels outwards
with positive velocities
The central object formed is a hot and lepton-rich protoneutron star (PNS).
The shock suffers continuously from the dissociation of
infalling heavy nuclei into free nucleons and light nuclei.
In addition as the shock crosses the neutrinospheres,
which relate to the neutrino energy and flavour dependent spheres
of the last scattering, additional electron captures release
a burst of electron-neutrinos.
This deleptonisation burst carries away energy of several \(10^{53}\)
erg/s (depending on the progenitor model) on a timescale of \(10-20\) ms.
This energy loss turns the expanding shock into a standing accretion shock
(SAS) with no significant matter outflow already about \(5\) ms after bounce.

It has long been investigated to revive the SAS via neutrino reactions,
leading to neutrino driven explosions (\citet{BetheWilson:1985}).
Unfortunately, there have been no explosions obtained
for progenitors more massive than \(8.8\) M\(_\odot\)
(\citet{Kitaura:etal:2006}, \citet{Fischer:etal:2009b})
in spherical symmetry.
A possible solution has been suggested and has been explored
only recently via multi-dimensional effects such as rotation,
convection and the development of known fluid instabilities
(see for example
\citet{Miller:etal:1993},
\citet{Herant:etal:1994},
\citet{Burrows:etal:1995}
and \citet{JankaMueller:1996}).
Such effects increase the neutrino
heating efficiency and help to understand
aspherical explosions,
see for example
\citet{Bruenn:etal:2006},
\citet{Janka:etal:2008a} and
\citet{MarekJanka:2009}.

\section{The model}

Our core collapse supernova model is originally based on 
Newtonian radiation hydrodynamics and
spectral three flavour Boltzmann neutrino transport,
developed by Mezzacappa \& Bruenn (1993a-c).
It has been extended to solve the general relativistic equations
for both, hydrodynamics and radiation transport,
as documented in Liebend\"orfer et al. (2001a,b).
Special emphasis has been devoted to accurately
conserve energy, momentum and lepton number
in \citet{Liebendoerfer:etal:2004}.
The following neutrino reactions are considered
\begin{eqnarray*}
e^- + p &\leftrightarrow& n +\nu_e, \\
e^+ + n &\leftrightarrow& p +\bar{\nu}_e, \\
e^- + \left<A,Z\right> &\leftrightarrow& \left<A,Z-1\right> +\nu_e, \\
\nu + N &\leftrightarrow& \nu + N, \\
\nu + \left<A,Z\right> &\leftrightarrow& \nu +\left<A,Z\right>, \\
\nu + e^\pm &\leftrightarrow& \nu + e^\pm, \\
e^- + e^+ &\leftrightarrow& \nu + \bar{\nu}, \\
N + N &\leftrightarrow& N + N + \nu + \bar{\nu}, \\
\nu_e + \bar{\nu}_e &\leftrightarrow& \nu_{\mu/\tau} + \bar{\nu}_{\mu/\tau},
\end{eqnarray*}
where \(N\in(n,p,\left<A,Z\right>)\) and
\(\nu\in(\nu_e,\nu_{\mu/\tau})\).
Nuclei are considered via a representative nucleus
with average atomic mass number \(A\) and charge \(Z\)
respectively.
The calculation of the reaction rates for these reactions
is based on \citet{Bruenn:1985}.
\(N\)-\(N\)-Bremsstrahlung has been implemented following
\citet{ThompsonBurrows:2001} based on \citet{HannestadRaffelt:1998}.
The annihilation of trapped electron neutrino pairs
was implemented recently and is documented in
\citet{Fischer:etal:2009a}.

The properties of (nuclear) matter
in core collapse supernovae are modelled via an EoS.
The two standard (hadronic) EoSs for matter
in nuclear statistical equilibrium (NSE) are from
\citet{LattimerSwesty:1991}
based on the compressible liquid drop model including surface effects
and the EoS from \citet{Shen:etal:1998a} based
on the RMF-approach and Thomas-Fermi approximation.
The first one can be applied using the three different
compressibilities \(180\), \(220\) and \(375\) MeV
and the low asymmetry energy of \(29.3\) MeV.
It is considered a rather soft EoS and was distributed to the
community as a subroutine and recently as a table as well.
The EoS contains contributions from electrons and positrons
as well as photons.
The latter EoS has a significantly larger asymmetry energy
of \(36.9\) MeV as well as a high compressibility of
\(281\) MeV which results in a stiff EoS.
It is distributed to the community as a baryon EoS-table.
For matter in non-NSE, formerly the approximation of
an ideal gas of Si-nuclei was used for the baryon EoS.
This leads to an increasingly inaccurate internal energy evolution,
especially in long term simulations of explosion models
where the explosion shock passes through the different
layers of composition given by the progenitor.
Recently, a nuclear reaction network has been implemented
(see \citet{Fischer:etal:2009b})
to be able to handle the nuclear composition
of the progenitor more accurately and
to include more mass of the progenitor
up to the hydrogen envelope (depending on the progenitor model).
Additionally, contributions from electrons and positrons as well as
photons and ion-ion-correlations (only for non-NSE)
are added based on
\citet{TimmesArnett:1999} and \citet{TimmesSwesty:2000}.

\begin{figure}[ht]
\centering
\includegraphics[angle=0,width=0.49\columnwidth]{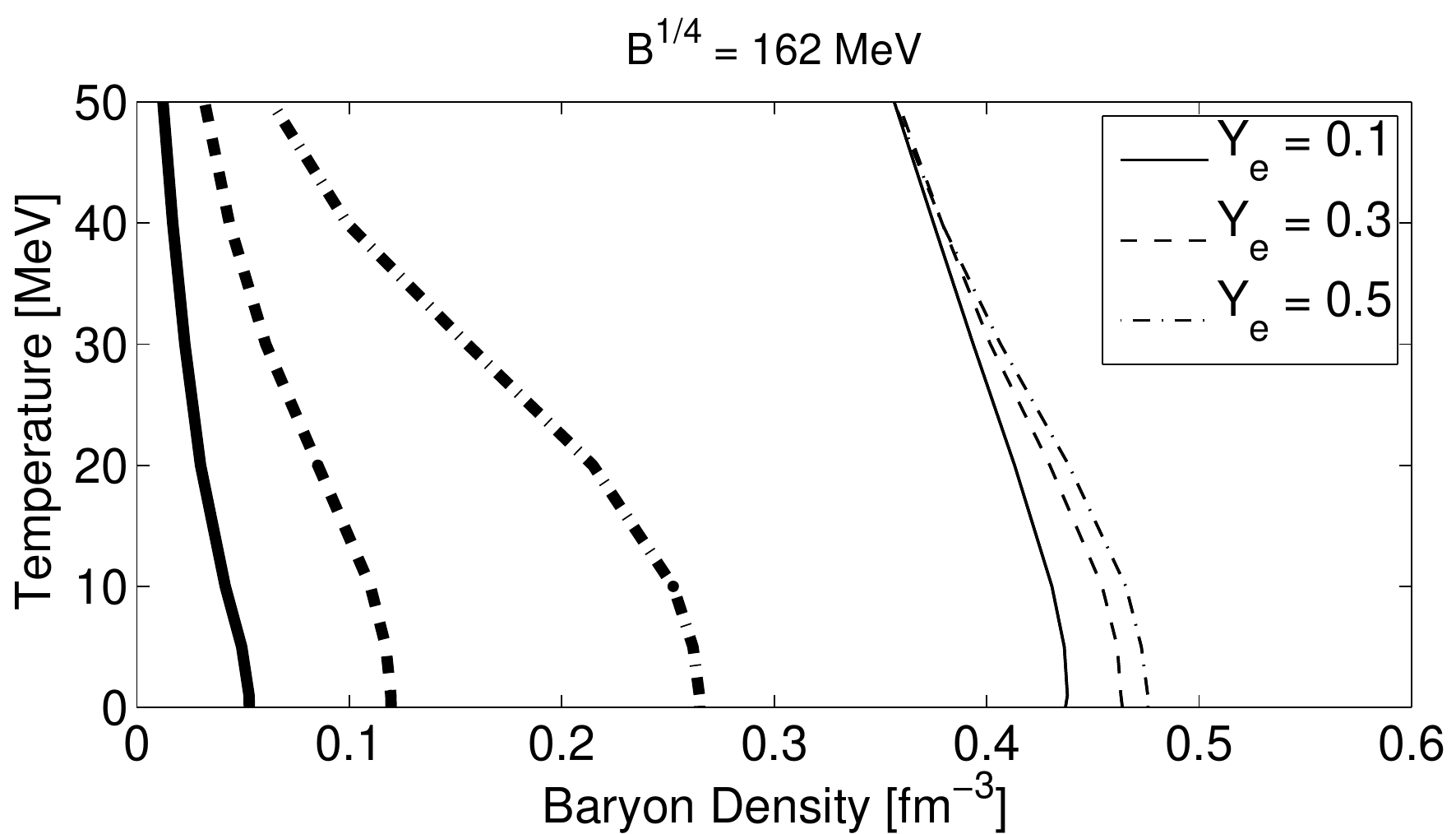}
\includegraphics[angle=0,width=0.49\columnwidth]{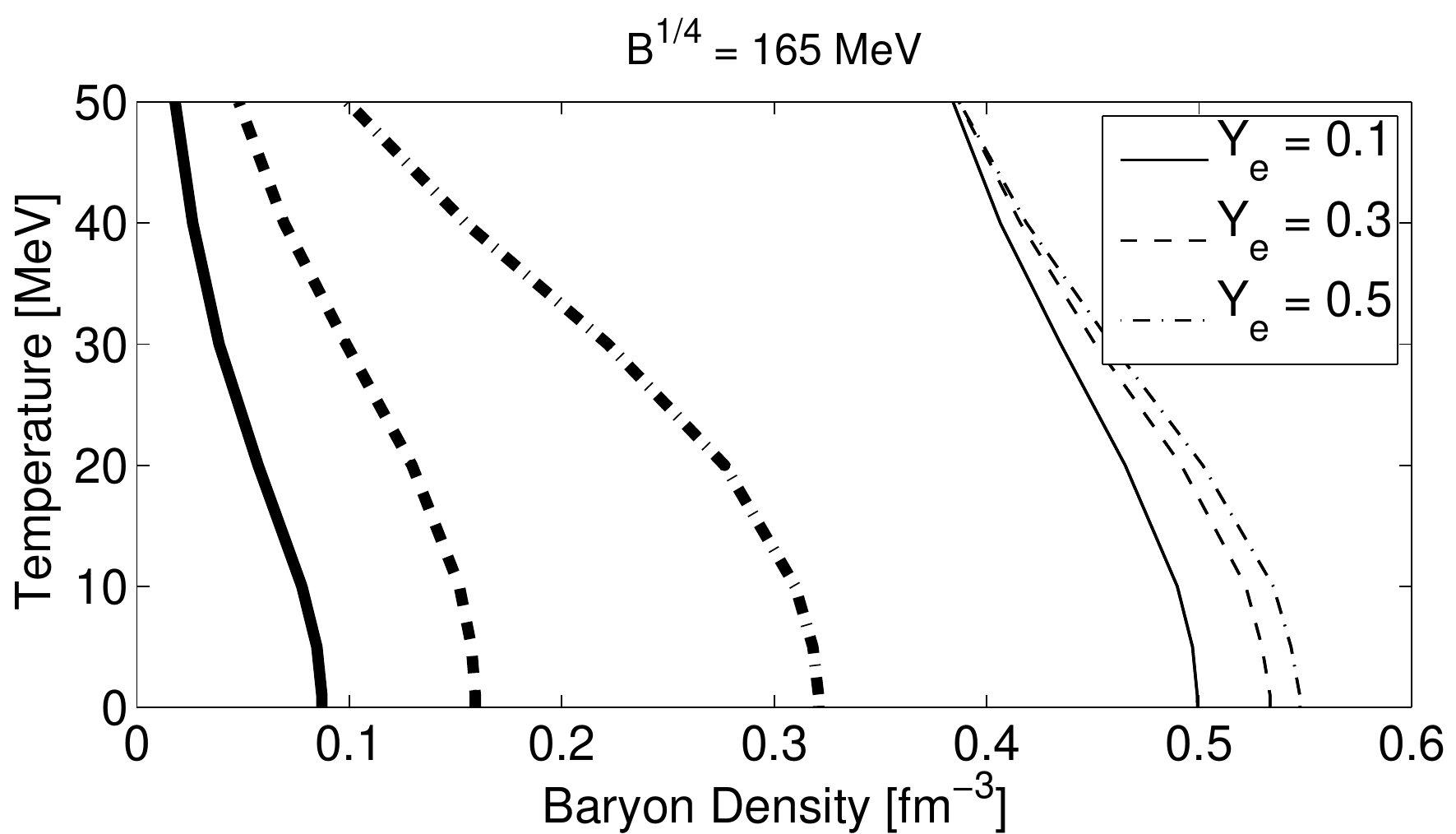}
\caption{Critical density as a function of the temperature
for different electron fractions, for the two quark EoSs
(left panel: \(B^{1/4}=162\) MeV, right panel: \(B^{1/4}=165\) MeV).
The lines show the onset (thick) and the endpoint (thin)
of the mixed phase from the QCD phase transition
from nuclear matter to quark matter.}
\label{fig:critdens}
\end{figure}
The quark EoS applied to the present investigation is based on the
widely applied MIT bag model.
The two values chosen for the bag constant are \(B^{1/4}=162\) MeV and
\(B^{1/4}=165\) MeV, which leads to a critical density close to nuclear
saturation density at low \(Y_e\) and finite temperatures and stable
maximum gravitational masses of \(M=1.56\) M\(_\odot\) and \(M=1.50\)
M\(_\odot\) respectively.
This is in agreement with the most precise NS mass measurement
of \(1.44\) M\(_\odot\) (Hulse-Taylor pulsar).
The EoS describes three flavour quark matter where up- and down-quarks
are considered to be massless and a strange-quark mass of \(100\) MeV
is assumed.
The behaviour of the critical density is illustrated in
Fig.~\ref{fig:critdens} for the two values of the Bag constant.
For high temperatures and low \(Y_e\), the critical density decreases.
This behaviour might change if a different quark matter description is applied.
The mixed phase is constructed using the Gibbs conditions, which leads
to an extended co-existence region and a smooth phase transition
where the entropy per baryon and the lepton number are conserved.

The nucleon and charge chemical potential as well as the nucleon mass
fractions are constructed from the quark chemical potentials and fractions,
which are then used to calculate the neutrino reaction rates and the transport
based on the hadronic description by \citet{Bruenn:1985}.
Since matter is opaque for neutrinos at densities where quark matter appears,
this approximation can be applied as long as the hydrodynamic timescale
is shorter than the diffusion timescale for neutrinos to diffuse out of the PNS.
This is the case for the post bounce scenario considered here,
where the hydrodynamic timescale is \(1-100\) ms and the diffusion
timescale is of the order seconds.
In addition, the timescale to establish \(\beta\)-equilibrium
is much shorter than the hydrodynamic and the diffusion timescales.
Hence, \(\beta\)-equilibrium is obtained instantaneously
for quark matter.

\section{The QCD phase transition during the early post bounce phase}

The appearance of quark matter in core collapse supernova
simulations is monitored using the quark matter volume
fraction \(x_Q\), defined as follows
\begin{equation*}
x_Q =
\left\{
\begin{array}{ll}
0 & \text{hadronic matter}, \\
]0,1[ & \text{mixed phase}, \\
1 & \text{quark phase}.
\end{array}
\right.
\end{equation*}
This is a standard procedure in nuclear physics
where a transition between two different phases
is constructed based on two different nuclear physics descriptions.
The conditions for the onset of the mixed phase as
illustrated in Fig.~\ref{fig:critdens} (thick lines)
are already reached at core bounce for the
\(10\), \(13\) and  \(15\) M\(_\odot\) progenitor models
from \citet{Woosley:etal:2002} under investigation.
However, the quark fraction reduces again after bounce
due to the density decrease in the expanding regime.
Only when the expanding bounce shock stalls and
turns into the SAS, the continued mass accretion causes
the central density to increase again.
The quark fraction rises on a timescale depending
on the mass accretion rate and is related to the
compression behaviour of the central PNS
given by the EoS.

The adiabatic index reduces for matter in the mixed phase.
Consequently, the smaller compressibility results in a higher
central density for matter in the mixed phase during the
post bounce accretion phase on a timescale of \(100\) ms.
The higher central density results in a different degeneracy
due to the different \(\beta\)-equilibrium, where
matter is found to be more neutron rich reducing
the electron fraction below \(Y_e=0.25\).
The additional loss of electron pressure support accelerates
the post bounce compression of the central PNS and favours even
higher central densities.

\begin{figure}[ht]
\centering
\includegraphics[angle=0,width=1.\columnwidth]{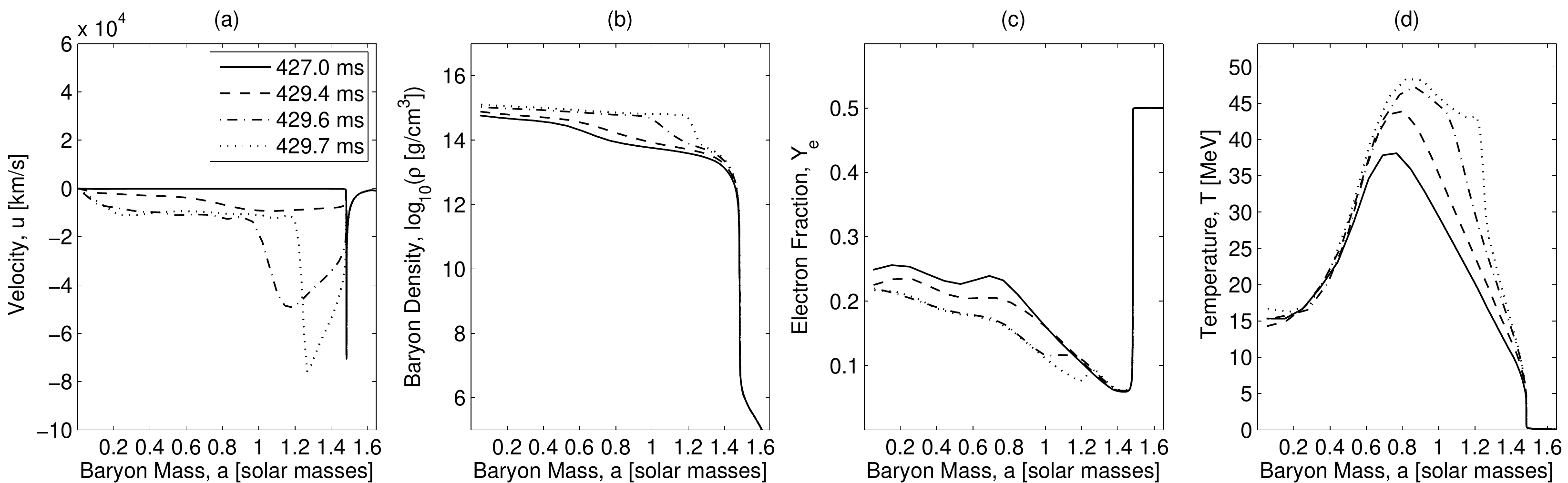}
\caption{Selected hydrodynamic variables
during the QCD phase transition
for several post bounce times.}
\label{fig:hydrostate-collapse}
\end{figure}
\begin{figure}[ht]
\centering
\includegraphics[angle=0,width=0.49\columnwidth]{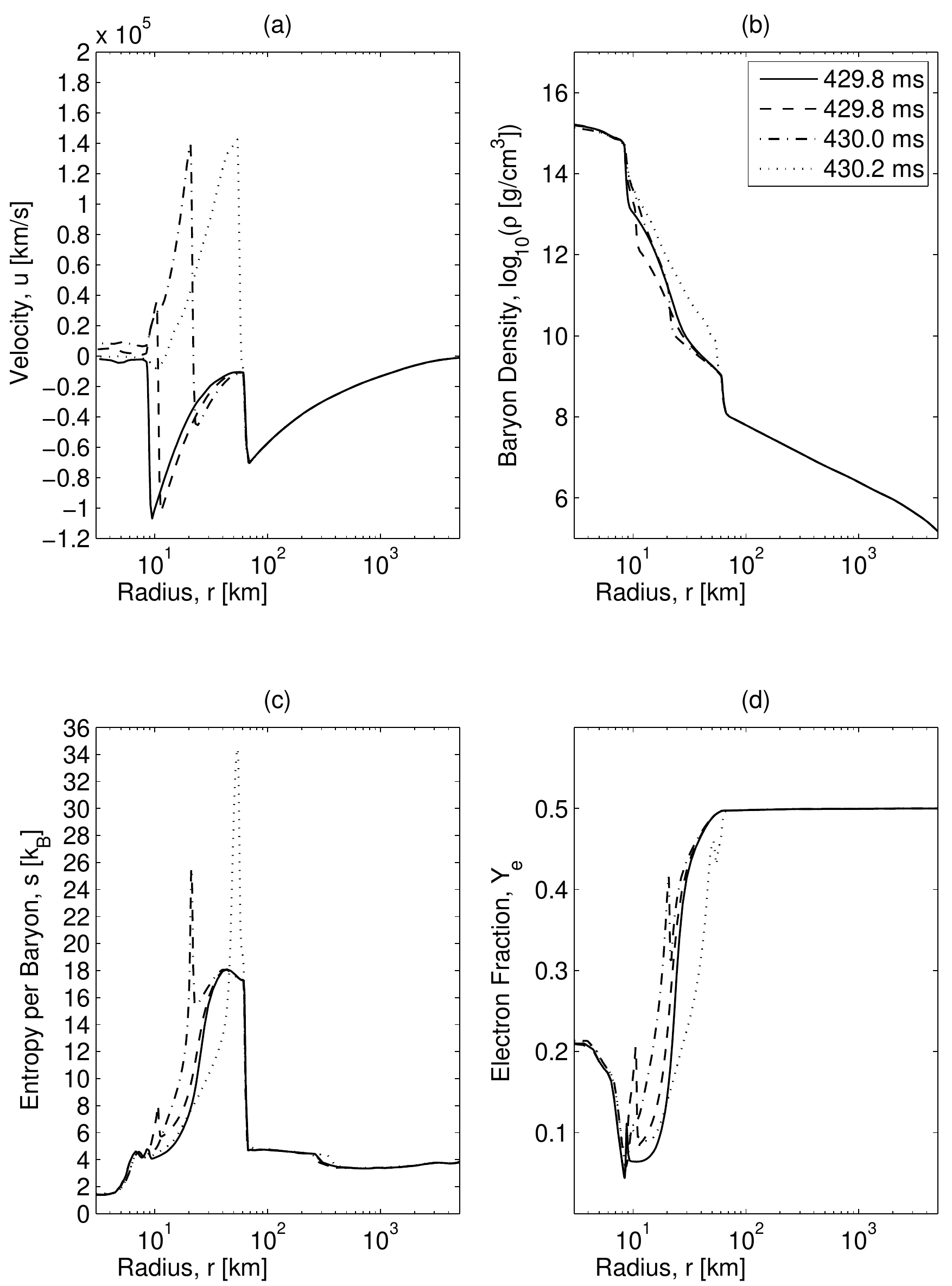}
\includegraphics[angle=0,width=0.49\columnwidth]{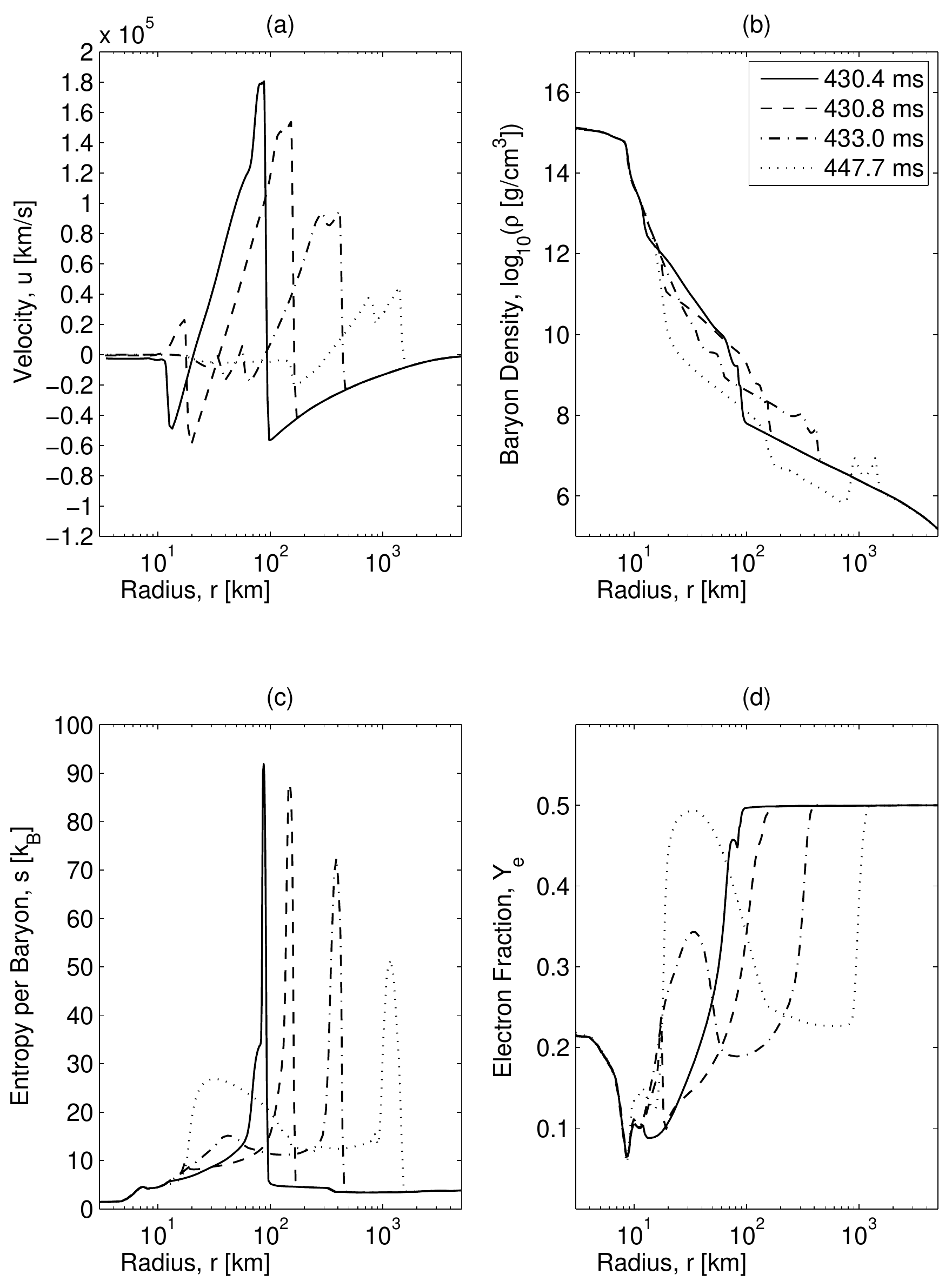}
\caption{Selected hydrodynamic variables
after the QCD phase transition
for several post bounce times,
illustrating the initial (left panel)
and the proceeding (right panel) explosion phases.}
\label{fig:hydrostate-expl}
\end{figure}
When a certain amount of mass of the PNS is converted into the mixed phase,
typically \(\sim 0.8\) M\(_\odot\), the reduced adiabatic index for
matter in the mixed phase causes the PNS to become gravitationally
unstable and contract.
The contraction is illustrated in Fig.~\ref{fig:hydrostate-collapse}
and proceeds into a supersonic adiabatic collapse.
Due to the contraction, density and temperature increase
which results in an increased degeneracy and hence
the electron fraction reduces
(see Fig.~\ref{fig:hydrostate-collapse} graphs (b) and (c)).
The collapse halts as a reasonable amount of matter
(depending on the progenitor)
inside the PNS is converted into the pure quark phase
via compression and the EoS stiffens again due to the
increased adiabatic index.
A stagnation wave forms which propagates outwards
and turns into a shock wave at the sonic point.
This scenario differs from the bounce
of the collapsing Fe-core, where the forming shock wave
travels outwards with positive velocities
immediately after its formation due to the
overshooting of the hydrostatic equilibrium configuration.
Here, the forming shock wave is not related to a
matter outflow and can be considered as a pure accretion shock
at the moment of its formation and shortly after.
This accretion shock propagates outwards because the thermal pressure
behind the shock is much larger then the ram pressure ahead of the shock.
The large thermal pressure behind the shock corresponds to the
converted hadrons into quarks since the shock spatially separates
mixed and hadronic phases.
As soon as the expanding accretion shock inside the PNS
reaches the PNS surface, it is accelerated along the
decreasing density and detaches from the mixed phase.
This behaviour as well as selected hydrodynamic variables
are illustrated in Fig.~\ref{fig:hydrostate-expl} (left panel).
This scenario is again different from the early post bounce behaviour
of the stalling bounce shock - since the bounce shock suffers from
the dissociation of infalling heavy nuclei, matter falling onto the
second shock is already dissociated and the nucleons are only
shifted to higher Fermi levels.
In this sense the second shock wave can accelerate quasi-freely.
Due to the large density decrease at the PNS surface over several
orders of magnitude from \(10^{15}\) to \(10^{9}\) g/cm\(^3\)
(see Fig.\ref{fig:hydrostate-expl} (b)),
the expanding shock wave reaches relativistic velocities of the
order \(~\sim 50\%\) the vacuum speed of light
(see Fig.\ref{fig:hydrostate-expl} (a)).
In other words, not only general relativistic effects are important
for the hydrodynamics equations as well as for the
neutrino transport in the presence of strong gravitational fields
inside the PNS but also kinetic relativistic effects become important
due to the high matter outflow velocities.

\begin{figure}[ht]
\centering
\includegraphics[angle=0,width=0.49\columnwidth]{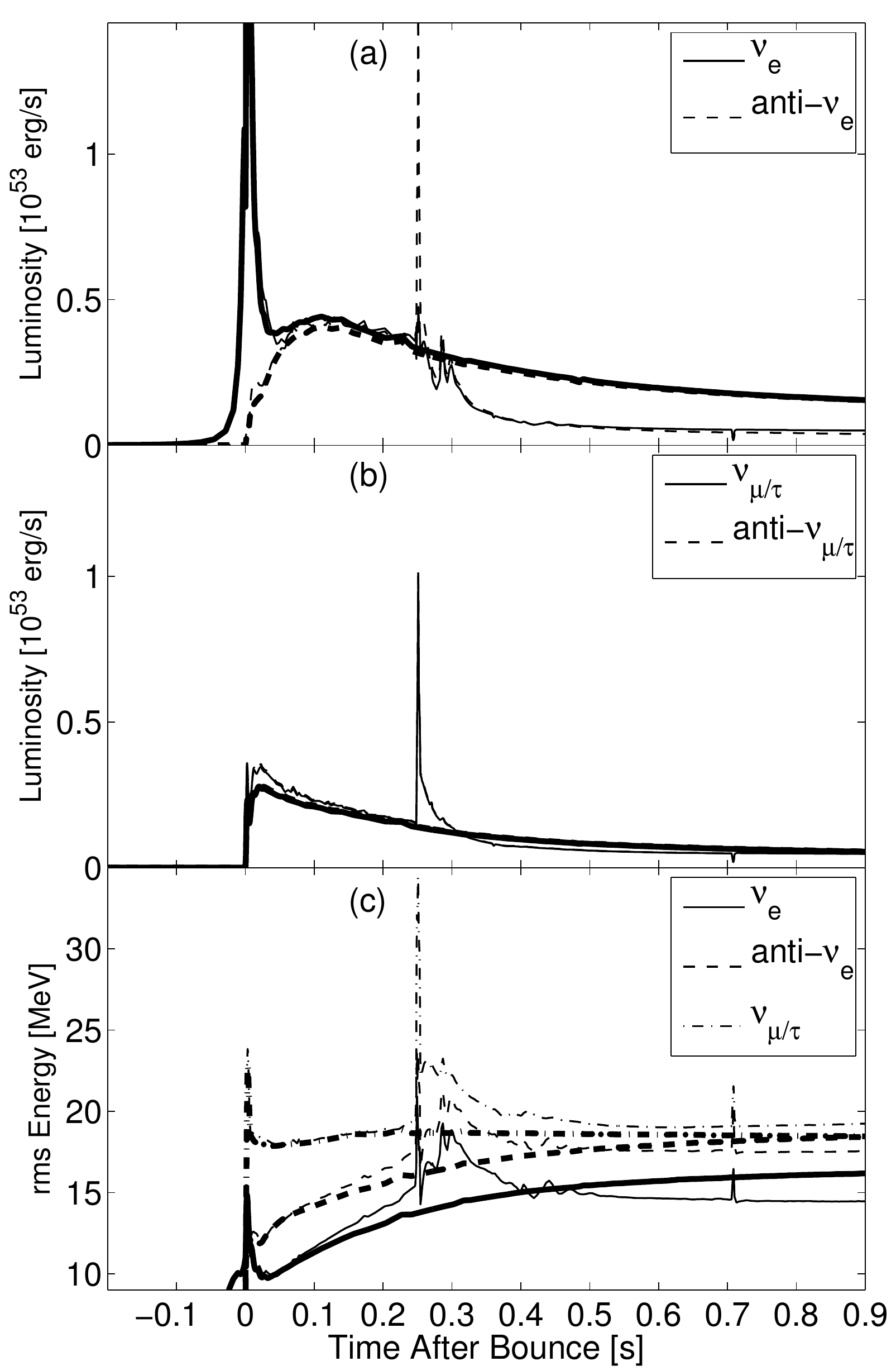}
\includegraphics[angle=0,width=0.49\columnwidth]{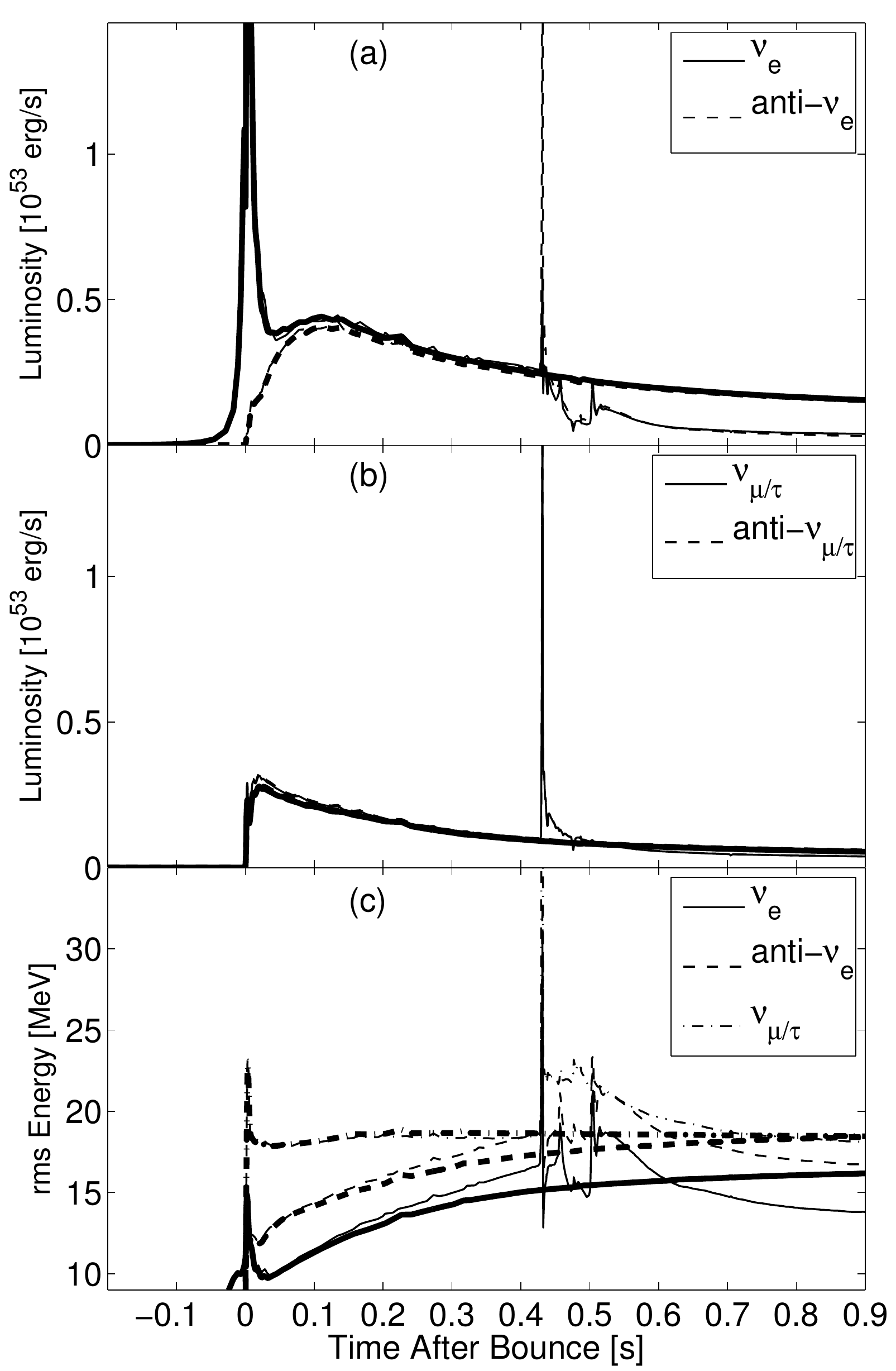}
\caption{Neutrino Luminosities and mean neutrino energies
for both quark EoSs (thin lines),
i.e. \(B^{1/4}=162\) MeV (left panel) and
\(B^{1/4}=165\) MeV (right panel),
in comparison to the pure hadronic EoS
from \citet{Shen:etal:1998a} (thick lines)
for the \(10\) M\(_\odot\)
progenitor model under investigation.}
\label{fig:lumin}
\end{figure}
The previously shock heated and highly deleptonised material
is accumulating onto the second expanding shock,
where it is shock heated again and
the degeneracy reduces such that \(\beta\)-equilibrium
is established at a larger value of the electron fraction
(see Fig.\ref{fig:hydrostate-expl} (d)).
This lifted degeneracy relates to a large fraction
of electron-antineutrinos that are emitted but still
can not escape because the matter conditions correspond
to the trapping regime.
Only when the second shock propagates across the neutrinospheres,
the additionally produced neutrinos can escape
which becomes observable in the neutrino spectrum
as a second neutrino burst.
It is accompanied by a significant increase of the
mean neutrino energies.
This burst is then dominated by electron-antineutrinos
due to the lifted degeneracy, again different from
the deleptonisation burst at bounce which is only
due to electron-neutrinos
(see Fig.~\ref{fig:lumin}).
The post bounce time of the second burst contains
correlated information about the contraction behaviour
of the central PNS, which in turn depends on the
(compressibility and asymmetry energy of the) EoS
for hadronic matter and the progenitor model,
as well as the critical conditions for the QCD
phase transition.
For the same progenitor model, a low critical density
(i.e. the small bag constant) corresponds to an
early phase transition.
The critical conditions for the quark matter phase transition
can be related to the delay of the second neutrino burst.
This delay depends on the central density increase
given by the mass accretion rate and the
hadronic EoS.

The expanding second shock wave finally merges with the first shock,
which remained unaffected by the happenings inside the PNS
at \(\sim 100\) km, and an explosion is obtained.
After the explosion has been launched,
the luminosities and mean neutrino energies decay.
In addition, a region of neutrino cooling develops
between the expanding explosion ejecta and the PNS surface
at the centre.
This can be seen in Fig.~\ref{fig:hydrostate-expl}
at the velocity profiles in the graphs (a) (right panel),
where initial mass inflow develops to
an additional weak accretion shock.
This expanding and contracting shock determines
the oscillating luminosity behaviour after
the second neutrino burst, illustrated in Fig.~\ref{fig:lumin}.
The following dynamical evolution of the explosion ejecta
can be approximated by an adiabatic expansion.
However the later appearance of the neutrino driven wind
and the related dynamic impact to the composition of the
ejecta will have to be analysed consistently in
explosive nucleosynthesis investigations.

\section{Summary}

\begin{table}[htpb]
\begin{center}
\caption{Selected properties of the PNSs
for the different progenitor models under investigation.
}
\begin{tabular}[b]{ c c c c c c c c }
\hline \hline
Prog. &  bag constant & \(t_{pb}\) & \(M_\text{PNS}\) & 
\(E_\text{expl}\) & \(\rho_\text{c}\) & \(T_\text{c}\) & \(Y_e\) \\
\([\)M\(_\odot]\)& \([\)MeV\(]\) & \([\)ms\(]\) &
\([\)M\(_\odot]\) & \([10^{51}\) erg\(]\) &
\([10^{14}\) g/cm\(^3]\) & \([\)MeV\(]\) & \\
\hline 
10 & 162 & 248 & 1.434 & 0.361 & 6.6069 & 13.14 & 0.2343 \\
10 & 165 & 429 & 1.482 & 1.080 & 5.8884 & 15.33 & 0.2488 \\
13 & 162 & 241 & 1.467 & 0.146 & 6.4565 & 13.32 & 0.2335 \\
13 & 165 & 431 & 1.498 & 0.323 & 5.6234 & 15.55 & 0.2462 \\
15 & 162 & 209 & 1.608 & 0.420 & 6.7608 & 14.10 & 0.2262 \\
15 & 165 & 330\(^1\) & 
1.700 & unknown\(^2\) & 5.4954 & 15.33 & 0.2479 \\
\hline
\end{tabular}
\end{center}
\small
\(^1\) time of black hole formation\\
\(^2\) black hole formation before positive explosion energies are achieved
\label{table:quark-runs}
\end{table}
The results of this first investigation of the
QCD phase transition in core collapse supernova simulations
of low and intermediate mass Fe-core progenitor stars
is summarised in Table~\ref{table:quark-runs}.
The simulations are launched from progenitor stars
of \(10\), \(13\) and \(15\) M\(_\odot\) from
\citet{Woosley:etal:2002}.
The post bounce times \(t_{pb}\) correspond to the
appearance of the second neutrino burst in the spectra
and the PNS masses \(M_\text{PNS}\) are taken
at the electron-neutrinospheres at some late post bounce
times when the simulations are stopped.
The central thermodynamic conditions, density \(\rho_\text{c}\),
temperature \(T_\text{c}\) and electron fraction \(Y_e\),
correspond to the initial PNS collapse.
A special model is the \(15\) M\(_\odot\) progenitor
using \(B^{1/4}=165\) MeV, where the maximum mass is reached
shortly after the QCD phase transition.
Hence, the PNS collapses to a black hole before the
already formed second shock is accelerated to
positive velocities.
Due to our co-moving coordinate choice, no stable solution
for the equations of energy and momentum
conservation can be obtained
and \(t_{pb}\) determines the time of black hole formation
after bounce.

The significant softening of the EoS for matter which is
in the mixed phase causes the PNS to collapse.
As a direct consequence of the softening of the EoS
for matter which reaches the pure quark phase,
a second shock wave forms.
The second shock accelerates and finally merges
with the first shock and launches an explosion.
The explosion energies \(E_\text{expl}\)
are give in Table~\ref{table:quark-runs}.
This mechanism was explored and found to even produce explosions
in spherical symmetry using general relativistic radiation hydrodynamics
based on spectral three-flavour Boltzmann neutrino transport,
where otherwise no explosion could be obtained.
The explosion energies in Table~\ref{table:quark-runs} are
evaluated at some late post bounce times
when the simulations are stopped
and might have not yet converged to the
final value of the explosion energy.
The simulations will have to be carried out longer.
However, moderate explosion energies of \(\simeq 10^{51}\) erg
could be found for the \(10\) M\(_\odot\) model using
\(B^{1/4}=165\) MeV (i.e. the later phase transition),
otherwise lower explosion energies are obtained.
The explosion energies might be shifted
to larger values when multi-dimensional effects
such as convection and rotation are taken into account.

The second shock forms in the high density regime
where neutrinos are fully trapped.
Hence, no direct observational identification
of the QCD phase transition can be found in the
emitted neutrino spectra.
It would be different if analysing the emitted
gravitational waves directly from the phase transition
(\citet{Abdikamolov:etal:2009}).
Unfortunately, gravitational waves have proven
difficult to detect.
Nevertheless, an indirect observation can be found in the
emitted neutrino spectra.
A second neutrino burst is released
when the second shock, formed during the PNS collapse,
propagates over the neutrinospheres.
This second burst is, due to the lifted degeneracy
of the shock heated hadronic material,
dominated by electron-antineutrinos.
The burst is accompanied by a significant increase of the
mean neutrino energies which might become resolvable
by neutrino detector facilities such as Super-Kamiokande
for a future Galactic core collapse supernova explosion
if quark matter appears.

One of the most important observations from supernova
explosions is the composition of the ejecta,
which is determined by explosive nucleosynthesis
investigations and are model dependent
(see for example Fr\"ohlich et al. (2006a-c)).
Of special interest is the production
of heavy elements for which rapid neutron captures
(\(r\)-process) in supernova explosion models
have long been investigated
(see for example
\citet{WoosleyBaron:1992},
\citet{Woosley:etal:1994},
\citet{Witti:etal:1994},
\citet{Otsuki:etal:2000},
\citet{Thompson:etal:2001},
Wanajo et~al.(2006a,b),
\citet{Arcones:etal:2007},
\citet{PanovJanka:2009}).
The abundances are calculated via post processing of mass trajectories
from explosion models and compared with the well known solar abundances.
Unfortunately, the very recent neutrino driven explosion models fail to
provide conditions favourable for the \(r\)-process, which are high
entropies per baryon, a fast expansion timescales and generally
neutron rich material.
Especially the latter aspect is a subject of research.
Although the explosion models achieved via the QCD phase transition
of the present article have to be analysed consistently with respect
to explosive nucleosynthesis, a reasonable amount of ejected matter
is found to be neutron-rich where \(Y_e\simeq0.35-0.45\).
In addition, moderate entropies per baryon and a fast expansion
timescale are obtained.
The conditions found may indeed be favourable for the \(r\)-process.

\section*{Acknowledgment}

The project was funded by the Swiss National Science Foundation
grant. no. PP002-106627/1 and 200020-122287 and
the Helmholtz Research School for Quark Matter Studies,
the Italian National Institute for Nuclear Physics,
the Graduate Program for Hadron and Ion Research (PG-HIR),
the Alliance Program of the Helmholtz Association (HA216/EMMI) and
the DFG through the Heidelberg Graduate School of Fundamental Physics.
The authors are additionally supported by CompStar, a research networking
program of the European Science Foundation, and the Scopes project
funded by the Swiss National Science Foundation grant. no. IB7320-110996/1.
A.Mezzacappa is supported at the Oak Ridge National Laboratory,
which is managed by UT-Battelle, LLC for the
U.S. Department of Energy under contract
DE-AC05-00OR22725.

\def\Discussion{
\setlength{\parskip}{0.3cm}\setlength{\parindent}{0.0cm}
     \bigskip\bigskip      {\Large {\bf Discussion}} \bigskip}
\def\speaker#1{{\bf #1:}\ }
\def\endDiscussion{}


\end{document}